\documentclass[
aps,
10pt,%
final,%
notitlepage,%
oneside,%
onecolumn,%
nobibnotes,%
nofootinbib,%
superscriptaddress,%
noshowpacs,%
centertags]%
{revtex4}

\usepackage{textcomp}
\usepackage{xcolor}
\usepackage{comment}
\usepackage{url}
\usepackage{multirow}
\usepackage{longtable}

\begin{document}

\title{Multifrequency study of GHz-peaked spectrum sources}

\author{\firstname{Yu.~V.}~\surname{Sotnikova}}
\affiliation{Special Astrophysical Observatory of RAS, Nizhnij Arkhyz, 369167 Russia}
\email[E-mail:]{lacerta999@gmail.com}

\author{\firstname{T.~V.}~\surname{Mufakharov}}
\affiliation{Shanghai Astronomical Observatory, 80 Nandan Road, Shanghai 200030, China}
\affiliation{Kazan Federal University, 18 Kremlyovskaya St., Kazan, 420008, Russia}

\author{\firstname{E.~K.}~\surname{Majorova}}
\affiliation{Special Astrophysical Observatory of RAS, Nizhnij Arkhyz, 369167 Russia}

\author{\firstname{M.~G.}~\surname{Mingaliev}}
\affiliation{Special Astrophysical Observatory of RAS, Nizhnij Arkhyz, 369167 Russia}
\affiliation{Kazan Federal University, 18 Kremlyovskaya St., Kazan, 420008, Russia}

\author{\firstname{R.~Yu.}~\surname{Udovitskiy}}
\affiliation{Special Astrophysical Observatory of RAS, Nizhnij Arkhyz, 369167 Russia}

\author{\firstname{N.~N.}~\surname{Bursov}}
\affiliation{Special Astrophysical Observatory of RAS, Nizhnij Arkhyz, 369167 Russia}

\author{\firstname{T.~A.}~\surname{Semenova}}
\affiliation{Special Astrophysical Observatory of RAS, Nizhnij Arkhyz, 369167 Russia}

\begin{abstract} 
Gigahertz-Peaked spectrum (GPS) sources are compact active galactic nuclei, presumably young precursors of bright radio sources. The study of GPS radio properties provides information about the features of synchrotron radiation in extragalactic sources. Also in applied research GPS sources are useful as compact stationary radio sources in the sky for astrometric purposes. This paper presents the results of a multifrequency GPS study based on quasi-simultaneous measurements with the RATAN-600 radio telescope during the 2006--2017. The catalog containing spectral flux densities measured at six frequencies (1.1, 2.3, 4.8, 7.7/8.2, 11.2, and 21.7 GHz) have been obtained. In addition, for the analysis of radio spectra, data from the following low-frequency surveys have been used: GLEAM (GaLactic and Extragalactic All-sky Murchison widefield array survey) and TGSS (Tata institute for fundamental research GMRT Sky Survey) and high-frequency measurements from Planck survey. A total number of 164 GPS and candidates to GPS have been identified (17 of them are new discoveries), which makes up a small fraction of GPS in the initial sample of bright AGNs, about 2\%. The physical properties and formation conditions of synchrotron radiation is found to be quite different in GPS of different AGN types. The deficit of distant GPS ($z > 2$) with low maximum frequencies (less than 1 GHz) is confirmed. The existing ``size -- peak frequency'' anticorrelation is continuous. The continuum radio spectra are found to become statistically steeper with increasing redshift.

\textit{\center{Keywords: {radio continuum: galaxies, galaxies: active, nuclei}}}

\end{abstract}
\maketitle

\textit{Published in Astrophysical Bulletin, Volume 74, Issue 4, pp.337-353, 2019}

\section{Introduction}

Gigahertz-peaked spectrum (GPS) sources are compact extragalactic sources, distinguished by their convex radio spectra with peak at several GHz (in the reference frame)\footnote{Hereafter, the spectral flux density is related to the frequency as $S_{\nu} \sim \nu^{\alpha}$}. They believed to be young predecessors of massive radio loud radio galaxies \cite{dea1,fanti90,fanti95}.
There are two classes of objects with properties similar to those of GPS, but they differ in peak frequency domain:
 Compact Steep Spectrum (CSS) have $\nu_{int}$<0.5 GHz
and High Frequency Peakers (HFP) have $\nu_{int}$> 5 GHz \cite{pea,bre,fan,dal}.
The above frequency boundaries are conventional, because the peak value may vary, and that is why in the literature these objects are often referred as GPS.
All of them are distinguished by high radio luminosity $L_{radio}\sim 10^{43-45}$ erg/s and compact size that less
than 1 kpc \cite{dea1} for GPS/HFP and about 20 kpc for CSS, low radio variability (a few percent) and small magnitude of polarization \cite{rudnick,dea1,dea2,tin1}.

GPS and CSS sources are believed to account for 10\% and 20\% of the brightest representatives of active galactic nuclei (AGN), respectively. However, recent studies with the RATAN-600 have revealed that the corresponding fractions are much smaller \cite{burs,mi1,mi2,mi22,mi}.
The shape of the spectrum after turnover frequency
is often explained in terms of the synchrotron self-absorption model (SSA -- synchrotron self-absortion) or with the free-free absorption model (FFA -- free-free absorption) \cite{snellen98}. The commonly adopted model explaining the small sizes of such objects is the youth scenario \cite{phillips82,carvalho85,wilkinson94,dea97,owsianik98,murgia03}. The presence of dense environment surrounding their central regions
can also hide their active and rapid expansion \cite{dea1}. The well-known ``linear size -- peak frequency'' anticorrelation \cite{fanti90} also indicates of the young age of GPS/CSS sources \cite{dea97}.

The origin of some GPS (about 10\%) is explained by the recurrent activity of radio galaxies \cite{baum90,pol}, when objects have a high radio luminosity for a long time. This hypothesis is confirmed by the discovery of diffuse radio-emission areas around some GPS, which may be the remnants of previous periods of activity \cite{baum90}. It was also suggested \cite{lee} that GPS quasars are blazars surrounded by a dense gas and dust medium. It can hide blazar properties during observations, despite the close line-of-sight location of the jet.

The observed radio spectra of GPS galaxies can be explained fairly well by projection effects \cite{snellen98}. There are few ready made scenarios for GPS quasars, and the samples are significantly ``contaminated'' by variable objects with temporarily inverted radio spectra  \cite{tor,tor1}. GPS quasars are the most compact (with sizes from several parsecs to several hundred parsecs) and have complex structures with a core and an emitting jet. Their compact size is due to the orientation of the jet, which is aligned close to the observers line of sight \cite{snellen98}.

GPS sources are obviously a heterogeneous group of extragalactic objects. The inclusion of a significant fraction of variable quasars in such samples affects the results of their analysis. An attempt of multidimensional ordering of the data about the various properties of GPS/HFP objects \cite{torniainen08} revealed many subgroups (clusters), which differ significantly in their properties. All known GPS quasars were reasonably included into Roma-BZCAT blazar catalog \cite{massaro09} . It has now been established that there are two types of GPS sources: classic, non-variable young objects, we will designate them as type 1 GPS, and variable objects often associated with blazars -- we will designate them as type 2 GPS. There is a slight difference in their spectra: the GPS of the first type have a narrower peak, whereas those of the second type usually has a wide peak \cite{planck2011,mi}. VLBI measurements are a reliable way to distinguish these objects \cite{bolton2006,vollmer2008}.

GPS sources are considered to be the predecessors of AGNs, as an early stages of their evolution \cite{fanti90,dea1,dal,tin1}. Hence that is interesting to study the possible connection of GPS with distant objects of the Universe \cite{falcke04,sadler15,coppejans16,coppejans2016,coppejans15,coppejans17}. The new measurements with high sensitivity at decimeter band  \cite{tgs,gle,callingham17} allowed the study of the compact objects with the peak in the spectrum below 1 GHz -- Megahertz Peaked Spectrum (MPS) objects. Today, MPS are key objects for an alternative search for extremely distant objects ($z > 6$) \cite{falcke04,coppejans2016,coppejans17}, when spectroscopic redshift measurements are difficult due to the multiple Ly-${\alpha}$ absorption lines on microwave background photons.

The aim of this work is to study the GPS radio properties over a wide frequency range (0.075-857 GHz) on time scales longer than 10 years. We analyze the properties of objects depending on the AGN class. Our analysis is based on measurements made with the RATAN-600 (1.1,2.3,4.8,7.7/8.2,11.2 and 21.7 GHz) during the 2006-2017 period of time. A significant amount of additional measurements consists of the data from the CATS astrophysical catalogs support system \cite{vo97,cats}. Additionally, the millimeter- and submillimiter-band flux densities were estimated for bright objects based on the analysis of the Planck survey map images. In our computations we adopted the following values of the cosmological constants: $\Omega_{m}$ = 0.27, $\Omega_{\Lambda}$ = 0.73, H = 71 (km/s)/Mpc.

\section{The sample properties and observations}

\begin{figure}[h]
%\onelinecaptionsfalse
\onelinecaptionstrue
\centerline{
\includegraphics[width=0.55\textwidth]{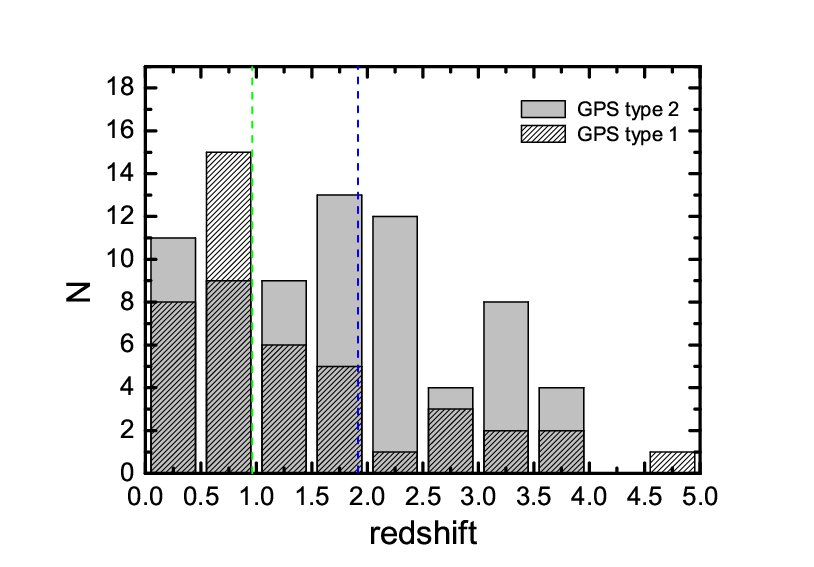}
}
\setcaptionmargin{0mm}
\captionstyle{normal}
\caption{Redshift distribution for the GPS objects of  type 1 and type 2. The dashed line indicates the median values of the redshift.}
\label{fig1}
\end{figure}

We study the complete sample of GPS source and candidates ($S_{5GHz} \geq 200$  mJy) from \cite{mi}. In that work a total number of 467 radio sources with a spectral maximum were selected, and 112 candidates were considered as GPS sources. GPS objects and candidate from this list were observed with the RATAN-600 within the framework of a planned research program in 2006-2017. Intermediate results of the monitoring for the 2006-2011 period were
published in \cite{mi1,mi2,mi}. The observations were carried out at the frequencies of 1.1, 2.3, 4.8, 7.7/8.2, 11.2, and 21.7 GHz. 
We selected GPS sources based on the criteria from \cite{kel,vris,dea1}, according to which the properties of the GPS radio spectrum should which should coincide with the theoretical spectrum of a  homogeneous self-absorbed synchrotron source with a power law electron energy distribution. These criteria includes: spectral indices of the optically thick and thin radiation 
regions is about $\alpha_{below}$=+0.5 and $\alpha_{above}$=-0.7, respectively \cite{vris}; the full width of the spectra at half maximum (FWHM) on the order of 1.2 frequency decades \cite{dea1,edwards}; and the variability index $V_{radio}\le25\%$. The redshifts values for objects studied are in a range from 0.01 up to 4.5 and are known for almost half of the sources (Fig.\ref{fig1}).

Spectral flux densities were measured with two sets of continuum radiometers, their parameters are presented in Table \ref{tab:sotn1}: (1) using secondary mirror $\textnumero$1 at the frequencies of 1.28, 2.25, 4.8, 7.7 / 8.2, 11.2, and 21.7 GHz (designated as ``1'' in the Table 1) and (2) with the three-frequency ``Eridan'' continuum radiometers using secondary mirror $\textnumero$2 (designated as ``2'' in the Table 1) at the frequencies of 4.8, 11.2, and 21.7 GHz.
The RATAN-600 continuum radiometers are the direct amplification receivers with a square-law detection.
All the radiometers were designed as the ``total power radiometer'' mode.
We use the standard data acquisition and controling system based on ER-DAS (Embedded Radiometric Data Acquisition
System) \cite{tsybulev}. The current level of the receiving equipment  is provided by low-noise uncooled amplifiers based on high-electron mobility transistors (HEMT) and digital signal processors in the data acquisition system.

\begin{small}
\begin{table}%[htbp]
\caption{RATAN-600 continuum radiometers.}
\label{tab:sotn1}
\centering
\begin{tabular}{rrllccrr}
\hline\hline
\multicolumn{2}{c}{$f$}  & \multicolumn{2}{c}{$\Delta f$} & \multicolumn{2}{c}{$\Delta S$}  & \multicolumn{2}{r}{FWHM} \\
\multicolumn{2}{c}{(GHz)} & \multicolumn{2}{c}{(GHz)}     & \multicolumn{2}{c}{(mJy/beam)}   & \multicolumn{2}{r}{(arcsec)} \\
\hline
      1     &     2           &   1        &  2              &  1   & 2 & 1 & 2        \\
\hline
\hline
$21.7$    & $21.7$ & $2.5$  & $2.5$  &  $50$  & $95$  &   11 & 16.5 \\
$11.2$    & $11.2$ & $1.4$  & $1.0$  &  $15$  & $30$  &   16 & 23  \\
$7.7/8.2$ & ...    & $1.0$  & ...    &  $10$  & ...   &   22 & ... \\
$4.8$     & $4.8$  & $0.6$  & $0.8$  &  $5$   & $10$  &   35 & 53  \\
$2.25$    & ...    & $0.08$ & ...    &  $40$  & ...   &   80 & ... \\
$1.28$    & ...    & $0.06$ & ...    &  $200$ & ...   &  110 & ... \\
\hline
\hline
\end{tabular}

%\tablefoot{}
Designations:
$f$~-- central frequency ,
$\Delta f$~-- bandwidth,
$\Delta S$~-- sensitivity per beam in RA,
FWHM~-- angular resolution in RA.
\end{table}
\end{small}

The observations were made using two- and three-mirror antenna configurations. The angular resolution of the radio telescope depends on the declination of a source being observed, and due to the knife-edge shape of the beam, it is three to four times lower in declination than in right ascension. 
The detection limit for the RATAN-600 is approximately 5 mJy (3 s integration time) under good conditions at the frequency of 4.8 GHz and at an average antenna elevation.
The data were processed using an automated system of the reduction developed for the RATAN-600 continuum observations \cite{udovit}, based on the modules of the standard FADPS (Flexible Astronomical Data Processing System) package \cite{vo2} and designed for a mass interactive processing of output of the RATAN data. Estimates of measurement errors and the flux density calibration procedure can be found in \cite{NCPMi,NCPM,NCPM2,mi1,mi2,udovit}. The information for secondary calibration standards is taken from \cite{perley,ott,baars,tabara}. The systematic uncertainty of the absolute flux density scale (3-10\% at different frequencies) has not been included in the flux density error.
The total error of measured flux is determined by the calibration error and the antenna temperature error:

\begin{equation}
\label{1}
\left(\frac{\sigma_s}{S_\nu}\right)^2 =\left(\frac{\sigma_c}{g_\nu(h)}\right)^2+\left(\frac{\sigma_t}{T_{ant,\nu}}\right)^2
\end{equation}
where 
$\sigma_s$ is the total flux error;
$\sigma_c$ -- flux density calibration error;
$T_{ant,\nu}$ -- antenna temperature at the frequency ${\nu}$;
$\sigma_t$ -- antenna temperature mesurement error $T_{ant,\nu}$;
$S_\nu$ -- spectral flux density;
$g_\nu(h)$ -- calibration factor, that takes into account atmospheric absorption and telescope effective area dependence on the elevation $h$.
The elevation of the antenna $h$=90$^{o}$-$\phi$+$\delta$, where $\delta$ -- source declination,
$\phi$=43$^{o}$.653 -- is the telescope site latitude.  The average spectral flux density measurement error for the sample is 15\%, 7\%, 6\%, 5\%, 7\%, and 12\% at 21.7, 11.2, 8.2/7.7, 4.8, 2.3, and 1.2 GHz, respectively.

The continuum radio spectra of GPS sources are presented in \textit{Supplementary materials}. Some of the spectra can be analyzed in the interactive catalog of RATAN-600 measurements of BL Lacs \cite{blcat}. The crosses with downward arrows show the estimated fluxes of the ``hot spots'' (positive responce) on Planck maps. The spectra were analyzed using spg module of FADPS package.

The results of GPS measurements during the 2011-2017 period are presented in \textit{Supplementary materials} and also at the site of Strasbourg Astronomical Data Center (CDS)\footnote{http://vizier.u-strasbg.fr/viz-bin/VizieR}. In this paper we just provide a fragment in Table 2: Column 1 -- object name according to the NVSS catalog; Column (2) -- the mean observational epoch, and Columns (3-8) -- the inferred spectral flux densities and their errors as measured at RATAN-600 frequencies (indicated in the upper part of the figure).

\begin{table*}[ht!]
\setcaptionmargin{0mm} \onelinecaptionstrue
\captionstyle{flushleft}
\caption{\label{tab:sotn2} The flux densities of GPS sources and candidates for several epochs (RATAN-600 observations 2006-2017).}
%\hline
\begin{small}
\begin{tabular}{|c|c|c|c|c|c|c|c|}
\hline
NVSS name  & JD & $S_{21.7}, \sigma$ & $S_{11.2}, \sigma$ & $S_{7.7/8.2}, \sigma$ & $S_{4.8}, \sigma$ & $S_{2.3}, \sigma$ & $S_{1/1.3}, \sigma$ \\
	   &    & $(Jy)$             &      $(Jy)$        &    $(Jy)$             &   $(Jy)$          &      $(Jy)$       &      $(Jy)$        \\
\hline
032957$+$275615 & 2457861 &  $0.196\pm0.03$ & $0.384\pm0.04$  & $0.488\pm0.04$ & $0.686\pm0.04$ & $1.027\pm0.1 $ & $1.206\pm0.10$ \\
034729$+$200453 & 2455696 &                 & $0.172\pm0.02$  & $0.259\pm0.02$ & $0.334\pm0.02$ &                &                \\
		& 2457861 &  $0.116\pm0.02$ & $0.175\pm0.02$  & $0.251\pm0.02$ & $0.349\pm0.02$ & $0.354\pm0.02$ & $0.404\pm0.04$ \\
040305$+$260001 & 2455696 &  $1.490\pm0.09$ & $1.894\pm0.05$  & $2.194\pm0.07$ & $2.460\pm0.06$ &                &                \\
		& 2456939 &  $0.815\pm0.10$ & $1.044\pm0.10$  & $1.448\pm0.10$ & $1.834\pm0.10$ &                &                \\
		& 2457030 &  $0.856\pm0.10$ & $1.191\pm0.10$  & $1.479\pm0.10$ & $1.663\pm0.10$ &                &                \\
		& 2457831 &  $0.609\pm0.10$ & $1.057\pm0.10$  & $1.306\pm0.10$ & $1.461\pm0.10$ & $1.521\pm0.02$ & $1.395\pm0.10$ \\
040922$+$121739 & 2456173 &  $0.193\pm0.02$ & $0.252\pm0.02$  & $0.297\pm0.02$ & $0.358\pm0.02$ &                &                \\
		& 2457860 &  $0.144\pm0.02$ & $0.137\pm0.01$  & $0.157\pm0.01$ & $0.186\pm0.01$ & $0.252\pm0.02$ & $0.370\pm0.04$ \\
		& 2457033 &  $0.383\pm0.05$ & $0.280\pm0.03$  &                & $0.218\pm0.01$ &                &                \\
		& 2457090 &  $0.179\pm0.02$ & $0.219\pm0.02$  &                & $0.303\pm0.02$ &                &                \\
		& 2457124 &                 & $0.272\pm0.03$  &                & $0.250\pm0.01$ &                &                \\
		& 2457214 &  $0.223\pm0.03$ & $0.235\pm0.02$  &                & $0.248\pm0.01$ &                &                \\
\hline
\end{tabular}
\end{small}
\end{table*}

Based on the GPS selection criteria (see [1, 12, 13, 15, 45]), we selected 164 GPS objects and candidate objects whose data we give in Table 3. Seventeen among them (marked with ``*'') are identified for the first time. We also marked with asterisks ``*'' 30 more objects, which we identified for the first time as GPS candidates in our earlier paper [15]. Table 3 provides the following data:\\
(1) -- Name of the object;\\
(2) -- CSS, GPS, or HFP, and candidate ``g'' for objects with unknown redshift;\\ (3)-(4) -- Optical type and redshift according to NED (NASA/IPAC Extragalactic Database);\\ (5)-(6) -- The $\nu_{obs}$ and $\nu_{int}$ values;\\
(7)-(8) -- The spectral indices $\alpha_{below}$; and $\alpha_{above}$; \\
(9) -- The spectral index $\alpha_{353-857}$;\\
(10) -- The FWHM of the spectrum in the units of frequency decades;\\
(11) -- Spectral flux density variability at 11.2 GHz (\%);\\
(12) -- Morphology [28, 59];\\
(13) -- AGN type according to the Roma-BZCAT. The asterisks next to the values of spectral index $\alpha_{353-857}$ in Column (9) indicate the sources located distances greater than $2.5^{'}$ (up to $3.5^{'}$) from hotspots in Planck maps. In these cases the hotspots are often rather extended and the sources are located at their periphery.

In general, the objects studied are bright radio sources with a median flux density value of several hundred mJy (see Table 4). The GPS objects of type 2 are significantly brighter at almost all frequencies (Fig. 3).

\begin{figure}[h]
\onelinecaptionsfalse
\centerline{
\includegraphics[width=\textwidth,clip]{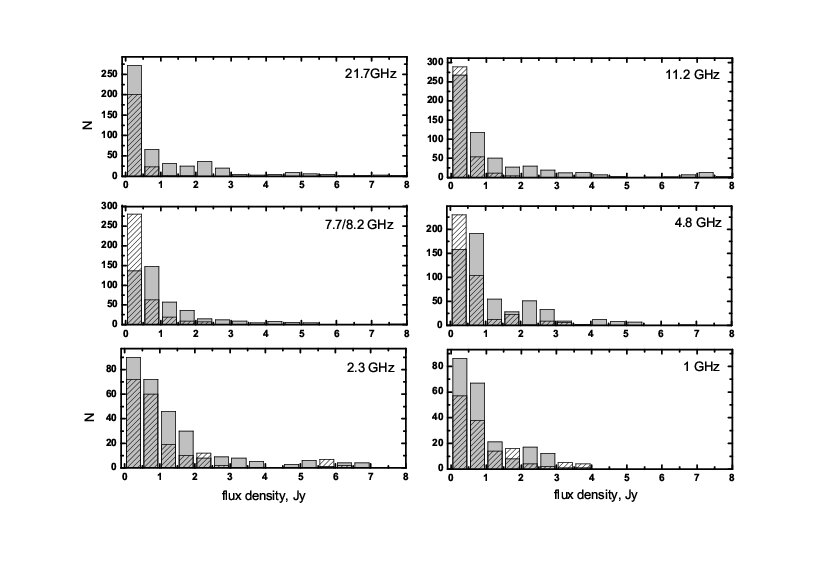}
}
\setcaptionmargin{0mm}
\captionstyle{normal}
\caption{The flux densities distribution at 1.2-21.7 GHz (RATAN-600) for GPS type 1 and 2 (hatched and gray, respectively). Objects with the flux densities exceeding 8 Jy were excluded for readability.}
\label{fig2}
\end{figure}

\begin{table*}[ht!]
\setcaptionmargin{0mm} \onelinecaptionstrue
\captionstyle{flushleft}
\caption{\label{tab:sotn3} The medians of the RATAN flux density values for GPS type 1 and 2.}
\begin{small}
\begin{tabular}{|l|c|c|c|c|c|c|}
\hline
 type   & $S_{21.7}, Jy$ & $S_{11.2}, Jy$ & $S_{8.2}, Jy$ & $S_{4.8}, Jy$ & $S_{2.3}, Jy$ & $S_{1.1}, Jy$ \\
\hline
	&        &       &      &      &      &      \\
type 1  &  0.19  &  0.21 & 0.28 & 0.41 & 0.66 & 0.62 \\
	&        &       &      &      &      &      \\
type 2  &  0.42  &  0.58 & 0.72 & 0.84 & 0.87 & 0.61 \\
	&        &       &      &      &      &      \\
\hline
\end{tabular}
\end{small}
\end{table*}

\clearpage
\newpage

\begin{table*}
\caption{\label{tab:sotn4} Parameters of GPS sources and candidates which were obtained with the RATAN-600 in 2006-2017 (part).}
\begin{small}
\begin{tabular}{|l|l|l|l|l|l|l|l|l|r|l|l|l|}
\hline
 $NVSS$ & sp.  & opt & $z$ & $\nu_{obs}$ & $\nu_{int}$ &~ $\alpha_{below}$, $\sigma$ &~ $\alpha_{above}$, $\sigma$& $\alpha_{353-857}$& $FWHM$ & $Var_{11.2}$ & morph. & AGN type \\
 name   & type & type &     &$GHz$    ~~  & $GHz$ &                                  &                                 &                   &        & \% &    &          \\
\hline
  (1)   & (2)      &~(3)  &~(4)  &(5)~~~~   & (6)   &~~~~~~(7)                  &~~~~~~(8)                       &~~~~(9)            &~~~~(10)   & ~~~(11) &~~~~~(12) &~~~~~(13) \\
\hline
000319$+$212944  & GPS & QSO & 0.45 & $ 2.3$ &  3.3 & 0.429$\pm$0.006 & -0.948 $\pm$0.013   & 2.7 $\pm$1.3  * & 1.5  & 3.0  & cso &  FSRQ     \\
000346$+$480703  &     & --  & --   & $ 1.9$ &  --  & 1.037$\pm$0.1   & -1.004 $\pm$0.023   &                 & 1.3  & --   & cd  &  -        \\
000520$+$052410  & HFP & QSO & 1.89 & $ 2.0$ &  5.7 & 0.189$\pm$0.027 & -0.473 $\pm$0.010   &                 & 1.8  & 25.0 & cso &  FSRQ     \\
000800$-$233917* & HFP & QSO & 1.41 & $ 3.4$ &  8.1 & 0.450$\pm$0.004 & -0.550 $\pm$0.007   &                 & 1.6  & 3.4  & -   &  FSRQ     \\ 
001610$-$001512  & GPS & --  & 1.57 & $ 0.9$ &  2.3 & 0.488$\pm$0.003 & -0.309 $\pm$0.002   &                 & 2.0  & --   & -   &  FSRQ     \\
\hline
\end{tabular}
\end{small}
\end{table*}

\section{The flux densities estimation at millimeter and submillimeter bands}

We derived the spectra of the sources in the millimeter- and submillimeter-bands based on WMAP\footnote{\tt http://lambda.gsfc.nasa.gov} (23-94 GHz)  \cite{wmap9yr_temp} and Planck (30-847 GHz) data \cite{planck_rev}. Planck measurements have three times higher resolution and 10-times higher sensitivity. The WMAP and Planck data partially or completely coincide in time with the RATAN-600 GPS monitoring program in 2006-2017.

For the faint objects, which have no Planck measurements we estimated the upper flux density limit using the technique proposed in \cite{vo}. We used the maps of the microwave emission components obtained by Planck space telescope\footnote{\tt http://www.rssd.esa.int/Planck/} \cite{planck_rev}, and the Planck source catalogs, which are available from the Planck Legacy Archive -- PLA\footnote{\tt http://www.sciops.esa.int/index.php?project=planck\&page=Planck\_Legacy\_Archive}.

There is a hypothesis about the increased likelihood of a positive response (``hot spots'') on microwave background maps if a radio galaxy is located there \cite{eid}. It was shown that a fairly large number of extragalactic point sources with different spectral properties are detected at the $4\sigma$ on Planck maps containing signals of both the frequency channels and the cleaned CMB \cite{vo,vo1}. We used Aladin application \cite{al1,al2} to identify the spots with the objects studied and to measure the distance from the source to the center of the spot. We selected spots with the centers located no farther than $2.5^{'}$ and $3^{'} - 5^{'}$ from the sources at frequencies above 100 GHz and below 70 GHz, respectively.

We used Source Extractor program (SExtractor)\footnote{\tt http://terapix.iap.fr/soft/sextractor} \cite{sex} for photometric measurements of the signals from spots and to determine the integrated brightness of spots on Planck maps. Then we converted the inferred brightness temperatures into fluxes using the calibration curves \cite{vo}, which relate flux densities of the sources (in Jy) with the microwave background tem- perature on Planck maps (in Kelvins).

In this work we followed the procedure for determining the calibration curves using objects from lists \cite{mi2,mi} with known Planck catalog densities as calibrators. To control the technique, we compare the fluxes so obtained with the Planck catalog fluxes for sources with such data. The mean ratio $\overline{r}$ of the fluxes determined using the above technique to the fluxes given in the Planck catalog and the standard deviation of this mean ratio for the calibrators \cite{mi2,mi} lie in the $\overline{r} = 0.97-1.05$ and $\sigma_{r} = 0.2- 0.4$ intervals depending on the frequency.

The obtained upper flux limits for spots in Planck maps located in the vicinity of GPS sources were also included into the spectra of the objects (Fig. 2, downward-pointing arrows). We computed the two frequency spectral indices and constructed their distributions. The histograms are shown in Fig. 4. The hatched areas present the distributions of spectral indices based on Planck measurements exclusively, and the white areas correspond to all data including
both direct measurements from tne Planck catalog and the estimated fluxes of ``hot'' spots with the coordinates matching those of the sources.

\begin{figure}[h]
\onelinecaptionsfalse
\centerline{
\includegraphics[width=\textwidth,clip]{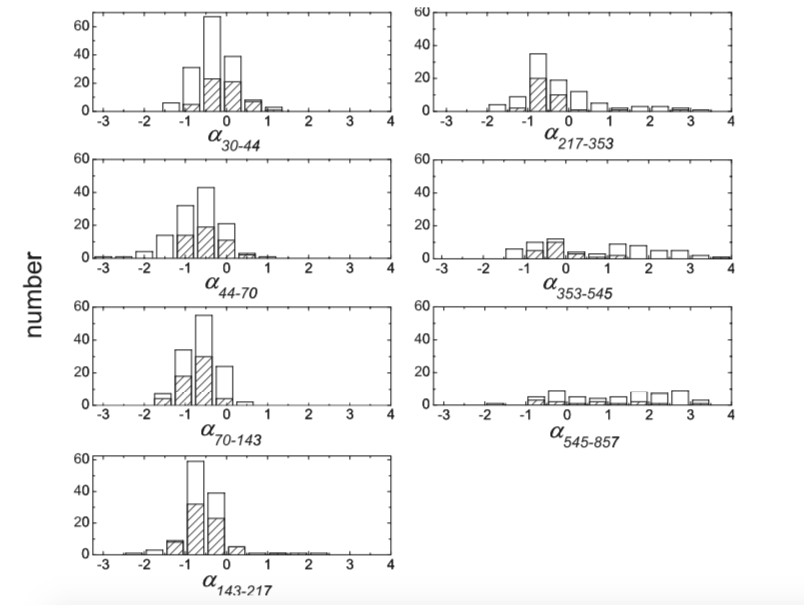} %sotnikova14.eps
}
\setcaptionmargin{0mm}
\captionstyle{normal}
\caption{The two frequency spectral indices distributions (from 30 up to 857 GHz). Spectral indices $\alpha$ were obtained from the Planck catalog fluxes measurements (hatched) and were computed based on entire data set including both Planck measurements and our upper fluxes limit estimates (white).}
\label{fig4}
\end{figure}

The distributions of the spectral indices $\alpha_{30-44}$, $\alpha_{44-70}$, $\alpha_{70-143}$ and $\alpha_{143-217}$, based on our upper flux estimates show the same pattern as the corre- sponding distributions based on Planck catalog data. This pattern also agrees well with the results based on two releases of the Planck catalog [67-69]. Thus the behavior of the spectral indices distributions reported in [67, 68] and our distributions (from $\alpha_{30-44}$ to $\alpha_{143-217}$) shows appreciable shift of the histograms toward lower $\alpha$, which is due to the steepening of the spectral indices of the sources. Such a distribution of indices at low frequecies is due to the predominance of sources with synchrotron radiation.

The histograms of the distributions of $\alpha_{217-353}$ are differ from those reported in [67, 68]. Most of the sources in our list, including those that have flux densities in the Planck catalog, have spectral indices in the $ -2 < \alpha_{217-353} < 1$ interval, and this fact also confirms the predominance of sources with synchrotron radiation. When going to frequencies above 217 GHz, a small fraction of the GPS objects shows a rise in the spectrum at 353-857 GHz, and the two frequency spectral indices $\alpha_{545-857}$ and $\alpha_{353-545}$ become greater than +1. According to the classification proposed in [69], such sources can be considered to belong to the ``intermediate synchrotron'' type. These are sources with signals detected at the frequencies of 545 GHz and 857 GHz and exhibiting both a strong synchrotron component and a significant dust component. Objects of this type were discovered, for example, when studying the spectra of extragalactic sources in [70].

Less than 10\% of the sources analyzed in [69] were classified as ``intermediate''. Their fraction increases substantially if we remove the condition of the completeness of the sample in terms of the high level of photometric noise in the signal and include the sources with lower fluxes.

If we consider only the distributions of two frequency spectra based on the Planck catalog data exclusively with a detection threshold of more than $4\sigma$ (Fig. 4, the shaded bars), when this sample consists practically of purely ``synchrotron'' objects. Of the 10 objects with the 857 GHz available in the Planck catalog, only four have indices $\alpha_{545-857} > 1$, which make up about 8\% of the sample. Lowering the detection threshold down to about $1-3\sigma$ increases the number of objects near which spots were found in Planck maps at the frequencies of 545 GHz and 857 GHz.

The rise of spectra at high frequencies can be due to the signal from cold Galactic dust present in Planck maps, although we cannot rule out the contribution of the proper internal dust of the sources. We cannot make more specific conclusions because the $\alpha_{353-857}$ spectral indices were inferred based on estimated data.

\section{Additional measurements}

Table 5 presents additional measurements from the CATS database in the form of list of catalogs. We mostly
used low-frequency and simultaneous mutifrequency measurements. We used low frequency measurements from GLEAM survey,
 which was conducted in 2013-2014 \cite{gle} at the 72-231 MHz
band at 20 frequencies with an angular resolution of $\sim 2{'}$ and with a flux density detection limit of 50 mJy. Our analysis
also includes TGSS survey measurements, which is conducted with the GMRT (Giant Metrewave Radio
Telescope) telescope at 150 MHz [39]. The first data release of the TGSS ADR1 (Alternative Data
Release) [39] is based on 2010-2012 observations. The catalog includes 620 thousand objects discovered with the flux detection 
limit of 7$\sigma$ and angular resolution  of $2{''}$. Both catalogs complete the objects spectral information at the frequencies  
 below 1.1 GHz. In addition, the analysis also includes the low-frequency measurements used in [13].
 
Additional low- and high-frequency measurements allowed us to determine a more reliable spectral classification, highlighting compact objects subclasses: CSS, GPS and HFP (Table 3).

\begin{table*}[ht]
\setcaptionmargin{0mm} \onelinecaptionstrue
\captionstyle{flushleft}
\caption{\label{tab:sotn5} Additional measurements (catalogs from the CATS database).}
\begin{small}
\begin{tabular}{l|l|l}
\hline
Frequency,        & Catalog & Reference \\
MHz                  &             &  \\
\hline
\hline
72-231                            & GLEAM   &  \cite{gle} \\
74                                & VLSS    &  \cite{vlss}      \\
150                               & TGSS    &  \cite{tgs}       \\
365                               & TXS     &  \cite{txs}      \\
318-750                           & Kuehr   &  \cite{ku81}     \\
80-2700                           & PKS90   &  \cite{pks90}    \\
325                               & WENSS   &  \cite{wenss}    \\
232                               & MIYUN   &  \cite{Zhang97}  \\
352                               & WISH    &  \cite{WISH}     \\
408                               & MRC     &  \cite{MRC}      \\
611                               & NAIC    &  \cite{NAIC}     \\
1365, 1665                        & GPSDa   &  \cite{GPSDa}    \\
1365, 1665, 2300                  & GPSTi   &  \cite{tin1}    \\
325, 608, 1380, 1630, 2300, 2695  & GPSSt   &  \cite{GPSSt}    \\
1400                              & NVSS    &  \cite{nvss}     \\
1400                              & QORG    &  \cite{QORG}     \\
1415, 2700                        & MSL     &  \cite{MSL}      \\
2300, 2700                        & PKSFL   &  \cite{PKSFL}    \\
2300                              & VCS     &  \cite{petrov05,petrov06,kovalev07}    \\
1000-21700                        & KOV97   &  \cite{KOV97}    \\
1100-21700                        & NCPMi   &  \cite{NCPMi}    \\
1100-21700                        &         &  \cite{NCPM,NCPM2}    \\
1100-21700                        & GPSRa   &  \cite{mi2}    \\
1000-21700                        & SRCAT   &  CATS  \\
960-21700                         & SRCKi   &  \cite{SRCKi}    \\
22000-94000                       & WMAP    &  \cite{wmap9yr_temp}  \\
30000-857000                      & PCCS1   &  \cite{planck_rev}    \\
30000-857000                      &         &  \cite{planck16}    \\
43000                             & VLAC    &  \cite{vlac}     \\
\hline
\end{tabular}
\end{small}
\end{table*}

\section{Analysis of GPS radio spectra}

As a result of the combined radio spectra analysis of sample objects we determined the following parameters:
the peak frequency $\nu_{obs}$ and $\nu_{int}$ in the observer?s and reference frames, respectively; the spectral
index $\alpha_{below}$ and $\alpha_{above}$ of the optically thick and optically thin emission region, the FWHM (in the units of decades of frequency) and the variability index $V_{radio}$ [13, 15].

Classification of the objects by type of their spectra, AGN type and morphology are given in the Table 6.
About 40\% of the objects are type 2 GPS sources [29] with a spectral peak mostly at frequencies of 
$\nu_{int}>5$ GHz (HFP). The information about VLBI morphology is available for one third of the
sources -- mostly compact symmetric objects  (cso),  core-jet (cj) objects,  compact double (cd) objects, and
featureless objects (un) that could not be resolved on a scale of several milliseconds.

\begin{table*}[ht]
\setcaptionmargin{0mm} \onelinecaptionstrue
\captionstyle{flushleft}
\caption{\label{tab:sotn6} General classification of the GPS sample.}
\begin{small}
\begin{tabular}{|c|c|c|c|c|c|c|c|}
\hline
Spectral type                           & N  & type 1 & type 2 & cso & cj & un & cd  \\
\hline
\hline
HFP                            & 71 & 14    & 57    & 10  & 6  & 11 & --  \\
$\nu_{int}>$ 5 GHz             &    &       &       &     &    &    &     \\
GPS                            & 40 & 27    & 13    & 8   & 3  & 1  & 7   \\
0.5$\ge$$\nu_{int}$$\ge$ 5 GHz &    &       &       &     &    &    &     \\
CSS                            & 2  & 2     & --    & --  & 1  & -- & --  \\
$\nu_{int}<$ 0.5 GHz           &    &       &       &     &    &    &     \\
undefined                      & 51 & 50    & 1     & 4   & 2  & 1  & 2   \\
\hline
Total                          &164 & 93    & 71    & 22  & 12 & 13 & 9   \\
\hline
\end{tabular}
\end{small}
\end{table*}

\subsection{Spectral indices}

The distrbutions of the spectral indices are shown on Fig.5, hatched for type 1 and
solid gray for type 2 GPS sources.
According to the values of spectral indices and the Kruskal-
Wallis criterion ($p<0.005$), the subsamples of the
GPS objects of these types do not belong to the same
distribution (Table \ref{tab:sotn7}). Type 2 GPS objects have
statistically flatter spectra in the optically thick and
thin emission regions, that in standart models is usually considered as the presence of 
additional compact synchrotron components \cite{planck2011}.
The distribution of the $\alpha_{below}$ and $\alpha_{above}$ for GPS type 1 has a wider shape
and may indicate the heterogeneity of the sample and presence of the various properties of the radiating
medium.

\begin{figure}[h]
\onelinecaptionsfalse
\centerline{
\includegraphics[angle=0,width=0.6\textwidth,clip]{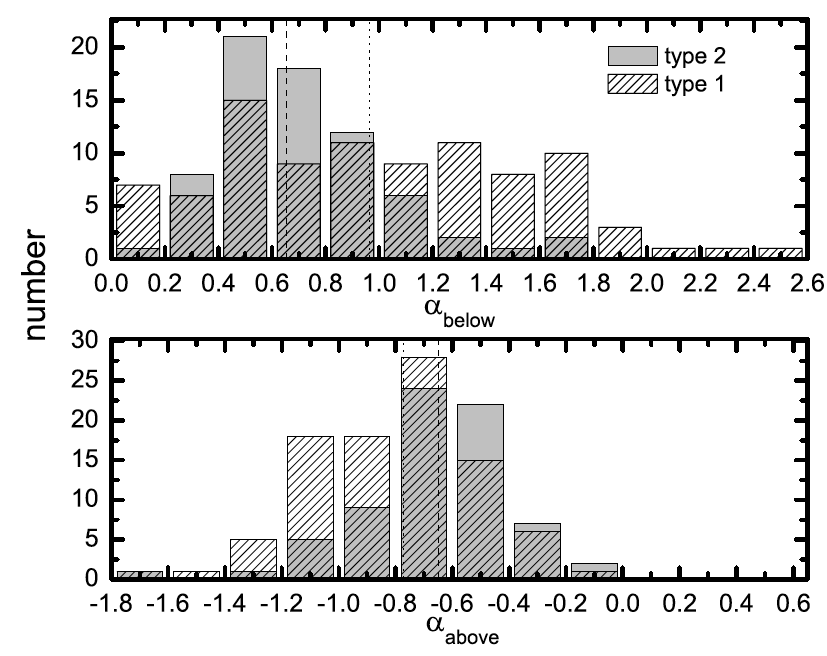}%sotnikova15.eps
}
\setcaptionmargin{0mm}
\captionstyle{normal}
\caption{
The distribution of high ($\alpha_{above}$) and low ($\alpha_{below}$)  frequency spectral
indices of the GPS objects. Statistically, second-type
GPS objects (solid gray) have flatter spectra both before and after the maximum.
The dotted and dashed lines show the median values of spectral indices for type 1 and type 2 GPS objects,
respectively.}
\label{fig5}
\end{figure}

\begin{table*}[ht]
\setcaptionmargin{0mm} \onelinecaptionstrue
\captionstyle{flushleft}
\caption{\label{tab:sotn7} Some parameters for GPS type 1 and type 2 (average values).}
\begin{small}
\begin{tabular}{|c|l|c|c|c|c|c|c|c|}
\hline
Type    & N   & z & $\alpha_{below}$ & $\alpha_{above}$ & $\nu_{int}$, & $FWHM$  & $\theta$, & $L_{5GHz}*10^{43}$,\\
	&     &   &                  &                  & GHz         &         & mas      & erg/s     \\
\hline
\hline
	 &     &     &             &             &           &     &           &          \\
 type 1  & 93  & 1.3 & +0.99 (0.5) & -0.80 (0.3) & 4.3 (0.6) & 1.4 & 2.2 (0.6) & 14 (3.2)\\
	 &     &     &             &             &           &     &           &          \\
\hline
	 &     &     &             &             &           &     &           &          \\
 type 2  & 71  & 1.8 & +0.71 (0.2) & -0.66 (0.2) & 14.1 (1.4)& 1.5 & 0.6 (0.2) & 56 (13.5)\\
	 &     &     &             &             &           &     &           &          \\
\hline
\end{tabular}
\end{small}
\end{table*}

There are statistical differences between the distributions of all measured parameters for type 1 and
type 2 GPS objects ($z$, $\nu_{obs}$, $\nu_{int}$, $FWHM$, angular size $\theta$,
and radio luminosity $L_{5GHz}$). The average value of 
$\alpha_{below}$ differs significantly from the theoretical limit for a uniform source of synchrotron
radiation, $\alpha_{below}$ = 2.5, and this fact has been confirmed by observations [92]. Only
five objects have measured spectral indices $\alpha_{below}$ of about 2 or greater (Table 8). 
This value is reliably determined for the galaxy $1447-34$ due to the large number of RATAN multifrequency measurements [13, 15] and low flux densities variability.

We determined 32 GPS objects with an ultra-steep radio spectra $\alpha_{above} < -1$ (Table 9).
There are no redshift measurements for 11 of them and five objects
have redshifts greater than 3. The sample contains a total of 17 objects withz > 3. 
We have not found a statistically significant trend in the relation $z$ vs. $\alpha_{above}$ for both GPS types
(Fig. 6). For comparison we have presented in the figure the measurements for 108 galaxies from [93]. These 
galaxies are candidates to ultra-steep spectrum objects with $\alpha_{150-5000 MHz} < -1$). The black
and gray dashed lines show direct approximations of the data. R\"ottgering et al. [93] found statistically 
significant correlation in the relation "$z$ -- $\alpha_{above}$", which supports the hypothesis that searching for objects
with ultra-steep spectra can be used to detect for high-redshift sources ($z > 3$).

We subdivided the initial sample into redshift bins of $\Delta z$=0.2 and determined the median $\alpha_{above}$ value
in each bin (Fig. 6b). The statistically significant correlation ($k_{sp}=-0.59$, $p<0.005$) have been determined, which
indicates the average spectral index steepening with increasing $z$ within the adjusted cosmological intervals.

\begin{table*}[ht]
\setcaptionmargin{0mm} \onelinecaptionstrue
\captionstyle{flushleft}
\caption{\label{tab:sotn8} List of objects with $\alpha_{below}$ values measured to be equal to 2 or greater.}
\begin{small}
\begin{tabular}{|l|l|r|c|}
\hline
Name     & $z$ & $\alpha_{above}$ & Optical type \\
\hline
\hline
0029$-$34 & --  & +2.1 (0.02) & --  \\
0806$-$29 & --  & +1.9 (0.01) & -- \\
1447$-$34 & 0.85& +2.5 (0.04) & G  \\
1845$+$35 & 0.76& +2.2 (0.04) & G  \\
2330$+$31 & --  & +1.8 (0.01) & --  \\
\hline
\hline
\end{tabular}
\end{small}
\end{table*}

\begin{figure}
\begin{tabular}{lr}
\begin{minipage}{0.5\linewidth}
\center{\includegraphics[width=\textwidth]{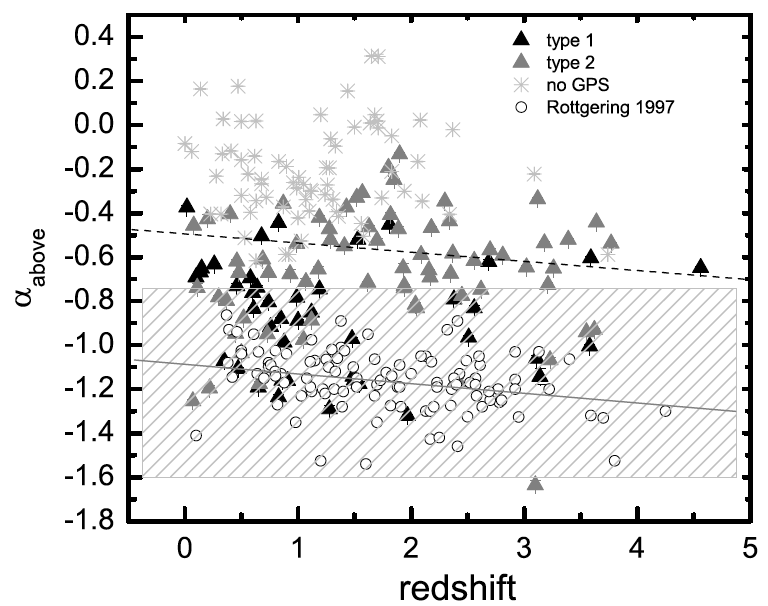}} \\ 
\end{minipage}
\begin{minipage}{0.5\linewidth}
\center{\includegraphics[width=\textwidth]{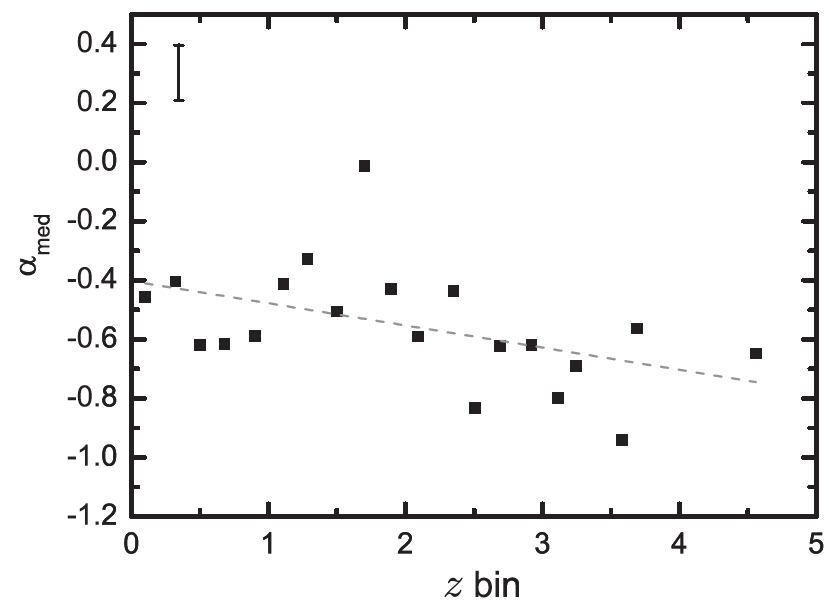}} \\  
\end{minipage}
\end{tabular}
\caption{Left: the ``$z$ -- $\alpha_{above}$'' relation for all objects of the initial sample (triangles and crosses); with additionally plotted galaxies
from [93] (circles); the dashed and solid lines show the linear interpolations of the formulas given in this paper and in [93]. Right: the same obtained by binnong 
with $\Delta z$=$0.2$; here $\alpha_{above}$ is computed as the median value in each redshift bin.}
\label{fig6}
\end{figure}

%\begin{figure}[h]
%\begin{minipage}[h]{0.9\linewidth}
%\centerline{\includegraphics[angle=0,width=0.85\textwidth,clip]{sotnikova16}} %16
%\end{minipage}
%\vfill
%\begin{minipage}[h]{0.9\linewidth}
%\centerline{\includegraphics[angle=0,width=0.80\textwidth,clip]{sotnikova17}} %17
%\end{minipage}
%\caption{Left: the ``$z$ -- $\alpha_{above}$'' relation for all objects of the initial sample (triangles and crosses); with additionally plotted galaxies
%from [93] (circles); the dashed and solid lines show the linear interpolations of the formulas given in this paper and in [93]. Right: the same obtained by binnong 
%with $\Delta z$=$0.2$; here $\alpha_{above}$ is computed as the median value in each redshift bin.}
%\label{fig6}
%\end{figure}

\begin{table*}[ht]
\setcaptionmargin{0mm} \onelinecaptionstrue
\captionstyle{flushleft}
\caption{\label{tab:sotn9}Objects with ultra-steep spectra $\alpha_{above}$<-1.}
\begin{small}
\begin{tabular}{|l|l|r|c|}
\hline
Name     & z & $\alpha_{above}$ & Optical/AGN type \\
\hline
\hline
0003$+$48 & --  & -1.0 (0.02) & --  \\
0048$+$06 & 3.58& -1.0 (0.01) & QSO \\
0108$-$12 & 1.54& -1.2 (0.01) & G  \\
0111$+$39 & 0.7 & -1.2 (0.01) & G/Blazar.un.type  \\
0204$+$09 & --  & -1.1 (0.01) & --  \\
0210$-$22 & 1.49& -1.1 (0.01) & G  \\
0318$+$16 & 0.91& -1.2 (0.01) & QSO \\
0557$+$24 & 3.2 & -1.1 (0.01) & FSRQ \\
0906$+$03 & 0.83& -1.2 (0.01) & G   \\
1009$+$06 & --  & -1.3 (0.01) & -- \\
1122$-$27 & 0.65& -1.2 (0.01) & -- \\
1227$+$36 & 1.97& -1.3 (0.01) & QSO \\
1237$+$20 & --  & -1.0 (0.01) & --  \\
1340$+$37 & 3.11& -1.1 (0.02) & QSO \\
1407$+$28 & 0.07& -1.3 (0.01) & QSO/BLac \\
1555$-$25 & --  & -1.5 (0.08) & -- \\
1600$-$00 & --  & -1.7 (0.04) & -- \\
1609$+$26 & 0.47& -1.1 (0.01) & G  \\
1753$+$27 & 0.86& -1.2 (0.01) & G  \\
1819$-$02 & --  & -1.0 (0.01) & -- \\
1826$+$27 & --  & -1.1 (0.01) & -- \\
1929$+$23 & --  & -1.4 (0.02) & -- \\
2022$+$61 & 0.2 & -1.2 (0.01) & FSRQ \\
2052$+$36 & 0.35& -1.1 (0.02) & G  \\
2131$-$12 & 0.5 & -1.1 (0.01) & FSRQ \\
2139$+$14 & 2.4 & -1.2 (0.02) & FSRQ \\
2143$+$33 & --  & -1.1 (0.02) & -- \\
2148$+$02 & --  & -1.0 (0.03) & -- \\
2208$+$18 & 3.14& -1.1 (0.04) & QSO \\
2237$-$25 & 1.28& -1.3 (0.02) & G \\
2316$-$33 & 3.1 & -1.6 (0.04) & QSO \\
2325$-$03 & 1.5 & -1.2 (0.01) & G  \\
\hline
\hline
\end{tabular}
\end{small}
\end{table*}

\subsection{$FWHM$ of the spectra}

One of the common GPS selection criteria is based on the FWHM (in decades of frequency). 
The authors of [1,45, 46] set it equal to 1.2 for the classical case.
However, the measured FWHM are almost always greater than 1.2 [13, 15]. We
obtained an average FWHM = 1.5 for type 2 GPS objects, which is somewhat greater (broader) than for
type 1 GPS objects (Fig. 7), which is equal to 1.4 decades of frequency.

The full width FWHM is related to the steepness of the spectrum and to the features of its flat part,
where $\alpha\sim0$. For some objects, this part is quite broad, as, e.g., for the blazar 2022$+$61, or, on the
contrary, very narrow (1340$+$37) (see the catalog of spectra). A broad spectrum is often associated
with the variability of radio emission; in this case we observe superpositions of several time-varying radio
spectra. Such spectra are more commonly found in blazars. The presence of several compact components
also plays a crucial role. Narrow spectrum is inherrent in the case of the lack of measurements.
To see whether the slopes of the spectra are consistent with the width, we analyzed the relation
between $\alpha_{below}$ and $\alpha_{above}$ and obtained a statistically significant linear trend (Fig. 8), with the regression coefficient
of $k=-0.4$ ($p<0.005$)
There is no dependence between spectral indices for the type 1 GPS sample under investigation.

\begin{figure}[h]
\onelinecaptionsfalse
\centerline{
\includegraphics[angle=0,width=0.55\textwidth,clip]{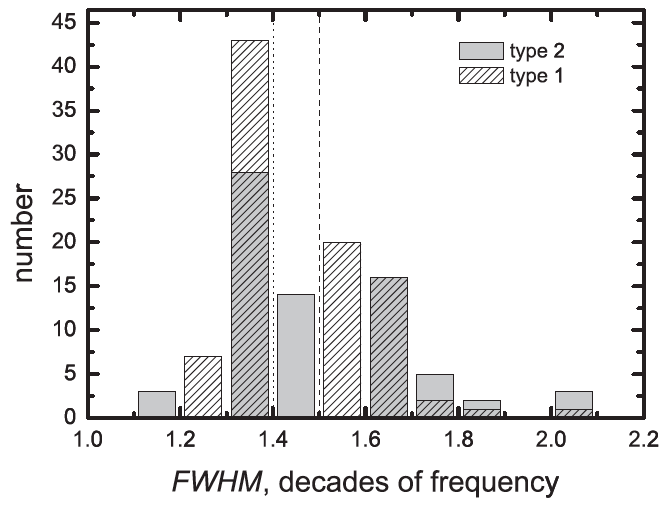} %18
}
\setcaptionmargin{0mm}
\captionstyle{normal}
\caption{The FWHM distribution for GPS objects of both types.
The dotted and dashed lines show the median FWHM values
for type 1 and type 2 GPS objects.}
\label{fig7}
\end{figure}

\begin{figure}[h!]
\onelinecaptionsfalse
\centerline{
\includegraphics[angle=0,width=0.7\textwidth,clip]{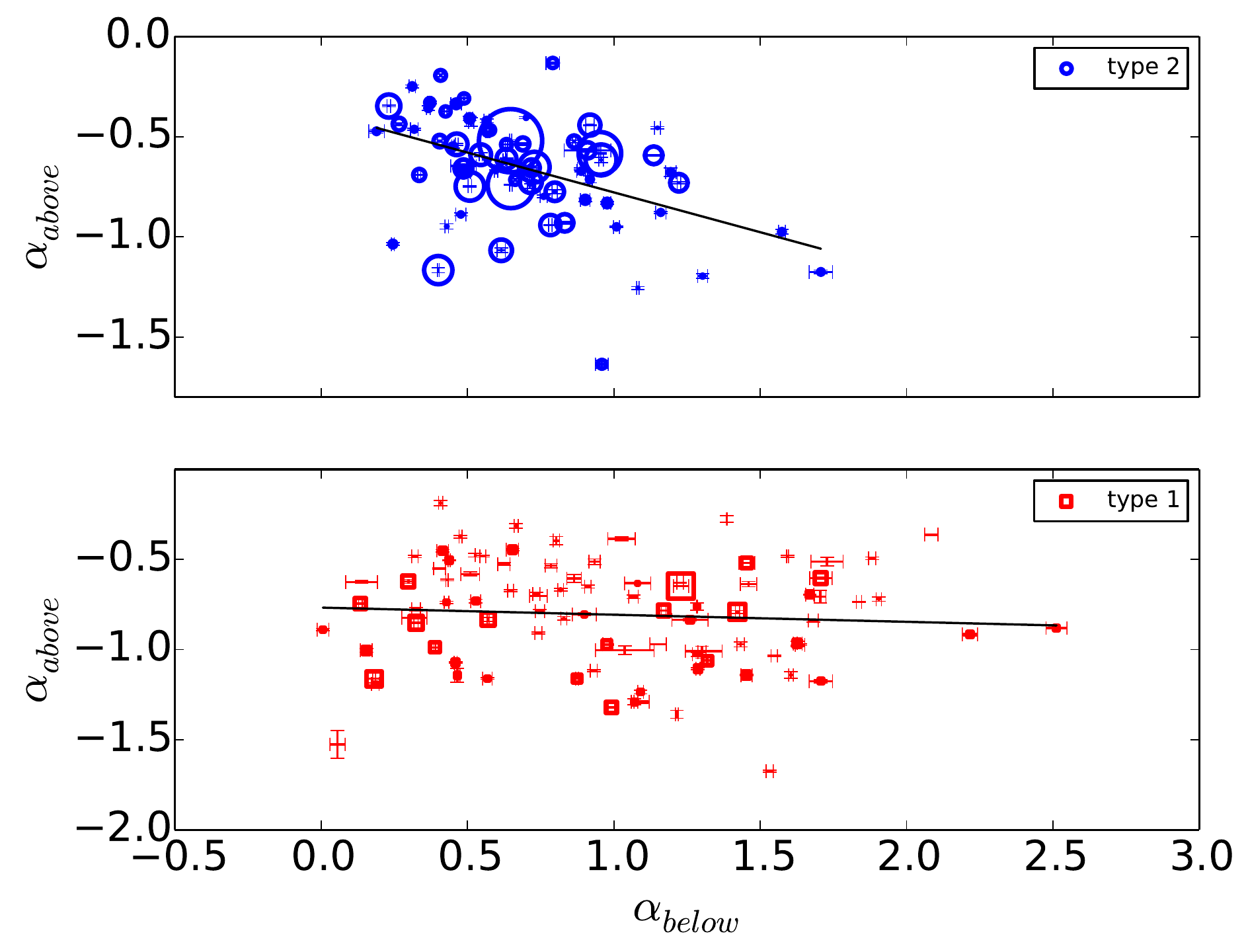} %19
}
\setcaptionmargin{0mm}
\captionstyle{normal}
\caption{Relation between the spectral indices $\alpha_{below}$ and $\alpha_{above}$
for GPS objects of both types. The sizes of the symbols denotes
the corresponding value of radio luminosity $L_{5 GHz}$.}
\label{fig8}
\end{figure}

\subsection{Estimation of linear size}
We determined the linear sizes of emitting regions by
the following formula \cite{slysh,kel}:

\begin{equation}
\label{2}
\nu_{max} = 8В^{1/5}S_{max}^{2/5}\theta^{-4/5}(1+z)^{1/5}
\end{equation}

it follows from:

\begin{equation}
\label{3}
\theta\approx1.345\frac{\sqrt{S_{max}}(1+z)^{1/4}}{\nu_{max}^{5/4}}
\end{equation}

where $B$ is the magnetic field strength in Gauss (we adopt the value of 100 $\mu$G for compact extragalactic sources with homogeneous distribution of magnetic field and relativistic particles \cite{mutel}), 
$S_{max}$ is the flux density value at the maximum of the radio spectrum in Jy; $\theta$ is the angular
size in mas, and $\nu_{max}$ is the observed frequency of the spectral peak in GHz.

The angular size estimates are made for a radio source with a homogeneous magnetic field and a power-law distribution of emitting particles with the self-absorption at the frequencies below $\nu_{obs}$. In the case where the object is a point
source for the beam pattern the registered emission is a sum of the emission of the source components.
Therefore formula (2) can be used to estimate the angular size of the upper limit of the emitting region.
On the whole, for the sample, angular size $\theta$ does not exceed 10 mas (Fig. 9). It is evident
from Fig. 10 that brighter objects of type 2 (with the average $L_{5GHz}\sim 56*10^{43}$  erg/s) have
more compact sizes (0.6 mas).

The peak frequency $\nu_{int}$ in the reference frame is
related to the peak frequency in the observer's 
frame as $\nu_{int} =\nu_{obs}(1+z)$. The additional measurements
used in this paper allowed us to extend
the range of $\nu_{int}$ values and analyze the ``$z - \nu_{int}$'' relation
(Fig. 11). Our result is in good agreement with earlier results, e.g., those reported
in [9, 45], extending the range of $\nu_{int}$ from 0.2 to 20 GHz.

At the redshifts of $z > 2$ a deficit of objects with the
$\nu_{int}$ less than 1 GHz is observed. Here the
solid line corresponds to the minimal value of $\nu_{int}$
in the sample at a certain redshift. For comparison,
we also show the $\nu_{int}$ values from the low-frequency (0.74-210 MHz) GLEAM survey [41]. Starting
with $z > 2$, our sample is dominated by HFP objects
($\nu_{int}$ > 5 GHz). The GLEAM sample (110 sources)
contains six sources for this redshift interval, with
$\nu_{int}$ spanning from 0.2 to 1 GHz. Fig. 12 shows the
well-known $\theta - \nu_{int}$ anticorrelation [2]. Also, these quantities are linearly related in logarithmic scale:

\begin{equation}
\label{3}
\log{\nu_{int}}=0.56(\pm0.03)-0.68(\pm0.01)*\log{\theta} 
\end{equation}

\begin{center}
or  $\nu_{int}\simeq\theta^{-0.68}$
\end{center}
\medskip

We found a significant anticorrelation for the entire sample ($k=-0.8, p<0.005$), which is strong for
type 1 GPS objects  ($k=-0.86, p<0.005$) and somewhat weaker for type 2 GPS objects ($k=-0.63, p<0.005$). The simple relation between the peak frequency
and angular size in Fig. 12 appears continuous for all intervals of $\nu_{int}$ and radio luminosity $L_{5GHz}$.

The existing anticorrelation often implies an evolutionary
connection between HFP, GPS, and CSS [2, 20].
Young bright compact objects have a spectral peak
at high radio frequencies; in the process of their
evolution they expand, the peak shifts toward lower
frequencies, and the luminosity decreases (``negative
luminosity evolution'' [96]). It is interesting to note the presence
of low-luminosity compact objects with a highfrequency
peak. This fact indicates that the initial
luminosity plays the crucial role in the morphological
evolution of extragalactic radio sources.

\begin{figure}[h]
\onelinecaptionsfalse
\centerline{
\includegraphics[angle=0,width=0.45\textwidth,clip]{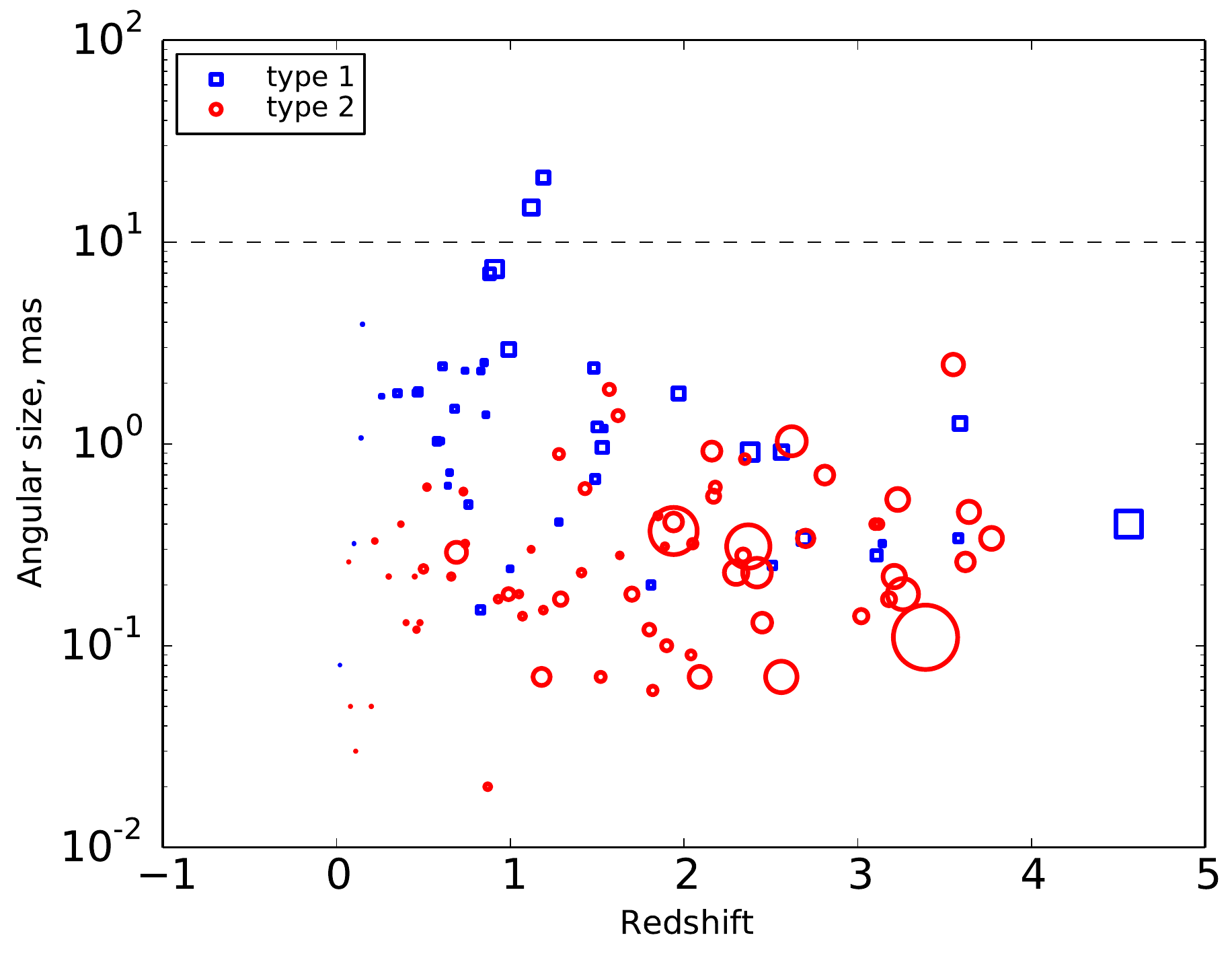} %20
}
\setcaptionmargin{0mm}
\captionstyle{normal}
\caption{Angular sizes $\theta$ of emitting regions for type 1 (squares)
and type 2 (circles) GPS objects at different redshifts. The sizes
of the symbols are proportional to the radio luminosity  $L_{5 GHz}$.}
\label{fig9}
\end{figure}

\begin{figure}[h]
\onelinecaptionsfalse
\centerline{
\includegraphics[angle=0,width=0.55\textwidth,clip]{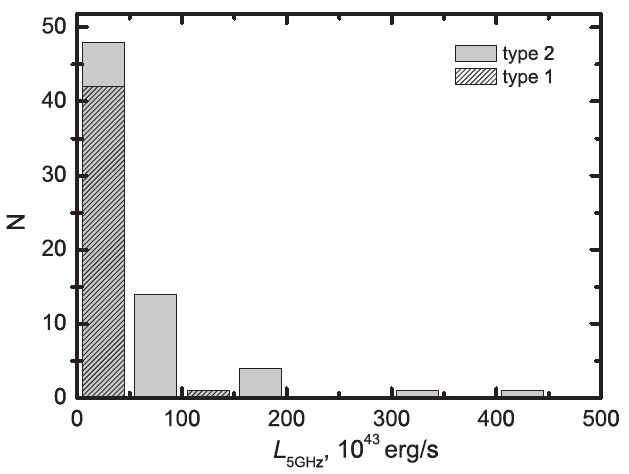} %21
}
\setcaptionmargin{0mm}
\captionstyle{normal}
\caption{The radio luminosities $L_{5GHz}$ distribution of type 1
(hatched) and type-2 (gray) GPS objects.}
\label{fig10}
\end{figure}

\begin{figure}[h]
\onelinecaptionsfalse
\centerline{
\includegraphics[angle=0,width=0.55\textwidth,clip]{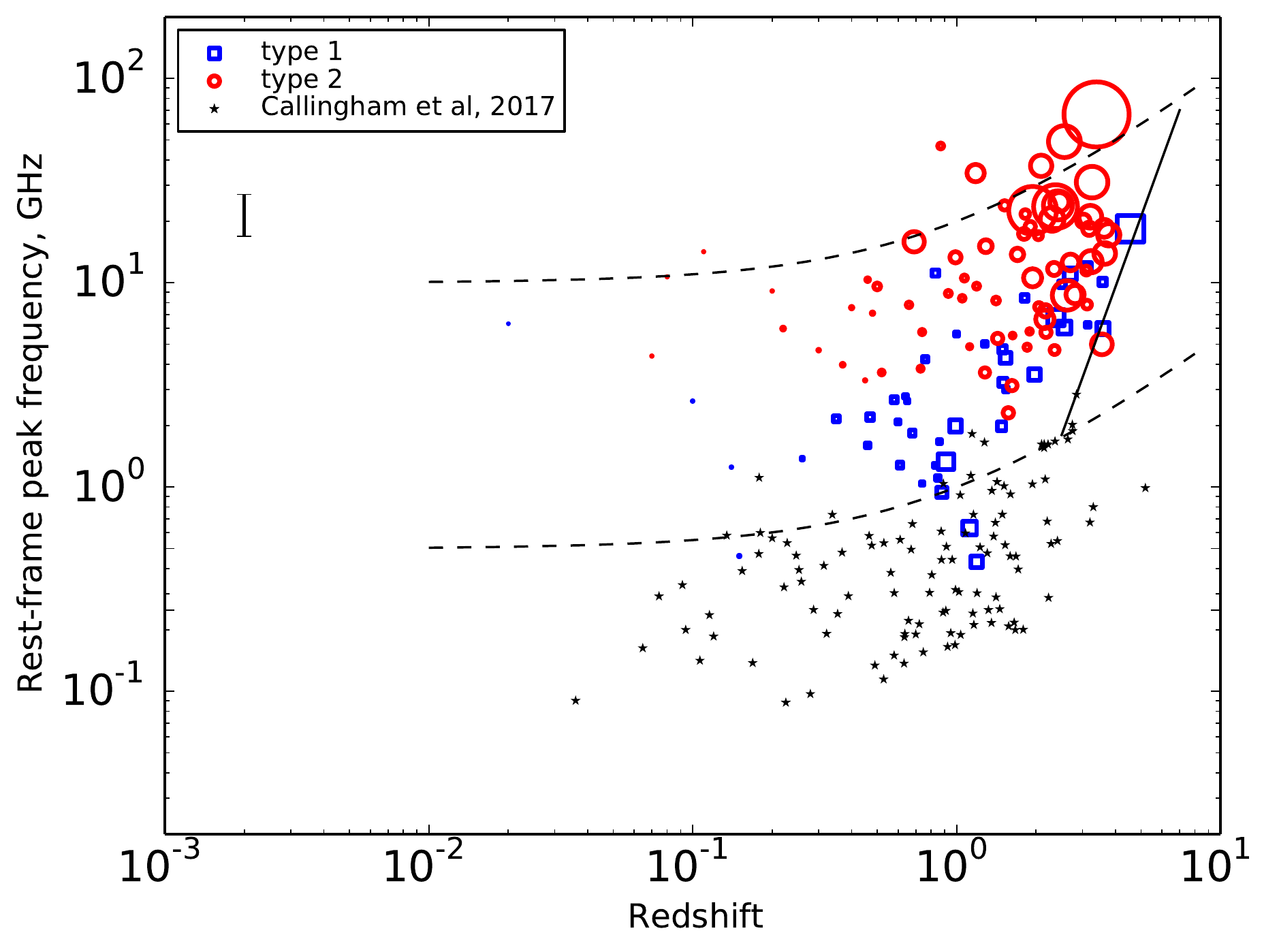} %22
}
\setcaptionmargin{0mm}
\captionstyle{normal}
\caption{Relation between redshift and peak frequency
in refernce $\nu_{int}$, for type 1 (squares) and type 2 (circles)
GPS objects. The symbol sizes correspond to the radio luminosity
$L_{5 GHz}$. The dashed lines show the evolutionary curves of the
frequency peak  for $\nu_{int}$ = 0.5 and 10 GHz; the solid line
corresponds to the minimum value of the  frequency $\nu_{int}$ at a certain redshift.
The black stars show the data for the comparison sample based on
objects of the GLEAM low-frequency survey [41].}
\label{fig11}
\end{figure}

\begin{figure}[h]
\onelinecaptionsfalse
\centerline{
\includegraphics[angle=0,width=0.55\textwidth,clip]{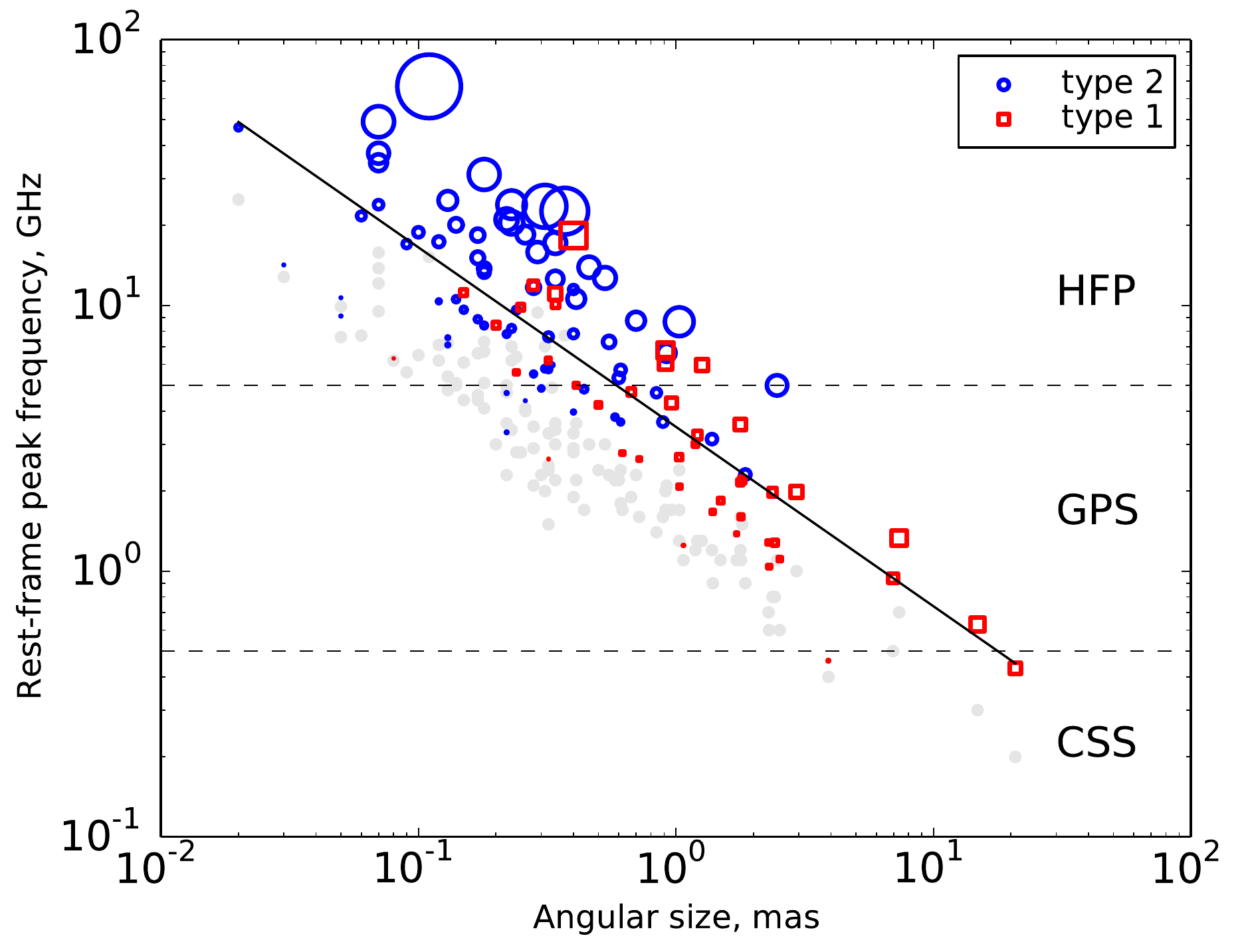} %23
}
\setcaptionmargin{0mm}
\captionstyle{normal}
\caption{Relation between the angular sizes of emitting regions
and the peak frequency in the reference frame, ``$\theta$ -- $\nu_{int}$'' , for type 1 (squares) and type 2 (circles) GPS objects. The gray circles
show the relation ``$\theta$ -- $\nu_{int}$'' for the entire sample. The black
line shows the linear regression of the anticorrelation in a logarithmic scale as $\nu_{int}\simeq\theta^{-0.68}$. The dashed lines
show the $\nu_{int}$ frequencies, which are equal to 0.5 and 1.0 GHz.
The sizes of the symbols correspond to the radio luminosity  $L_{5 GHz}$.}
\label{fig12}
\end{figure}

\section{Discussion}
A multifrequency study of a complete sample of GPS objects at radio wavelengths revealed that this type of AGN is
a heterogeneous group of compact extragalactic objects \cite{planck2011}.
On the whole, they can be subdivided into two large groups. The first group is associated with classical GPS representatives and includes young objects with low variability. The second group is associated with beamed jet oblects, mostly blazars \cite{planck2011}.
In this work we find that type 1 GPS objects are much rarer than it was previously assumed \cite{dea1,dea2},
and their fraction in the bright radio sources sample is less than 2\%. 
The sample we studied contains GPS objects of first ("classic") and second ("blazar") type in the proportion of 60\% to 40\%. 

Simultaneous multifrequency measurements of GPS objects performed with the same instrument over a long time period allowed us to eliminate the effects caused by variability of the objects.
Almost one third of the objects of the initial sample exhibited activity ($Var > 35{\%}$) at the radio frequencies over long time scales.
Many objects demonstrate the relatively quiet behavior in the radio domain.
Strong flux variations exceeding level of 50\% found in the form of rare irregular bursts. 
The GPS objects we studied are quite bright radio sources with the average radio luminosity 
of $10^{43-44} erg/s$ at 5 GHz. For the GPS of type 1 an average luminosity is $1.4*10^{44} erg/s$
and for type 2 -- $5.6*10^{44} erg/s$.

Statistical differences in the distribution of spectral indices for two types of GPS objects revealed a variety of ambient conditions for the propagation of radio waves. In general, type 2 GPS objects have narrower distributions of all spectral parameters, making up more homogeneous sample with common influence of relativistic effects.
%\textbf{Why it deleted in english?? --->
%В целом, GPS второго типа имеют более узкие распределения всех спектральных параметров,
%образуя более однородную выборку с общим
%влиянием релятивистских эффектов.
%Для GPS первого типа характер распределение спекральных величин
%выявляет выборку неоднородных по морфологии объектов.
%Результат согласуеся с работами \cite{torniainen08,vera2017},
%когда алгоритмы кластеризации свойств GPS не
%выявили их определенную морфологическую классификацию.}

For most GPS objects the spectral index of the optically thick emission region $\alpha_{below}$ is far from the theoretical limit of 2.5 \cite{Pacholczyk}:
the average values for type 1 and type 2 GPS objects reaches the value of +0.99 and +0.71, respectively. Three sources were found to have the spectral index of the optically thick emission region $\alpha_{below} \geq 2$: $J1447?34$ (2.5), $J0029?34$ (2.1), and $J1845+35$ (2.2).
The obtained relation ``$\theta$ - $\nu_{int}$'' in a good agreement with the results of a number of studies, e.g.,  \cite{fanti90,dea2,orienti14}, where the relation between the size and the internal peak frequency is found to be $\nu_{int}\simeq\theta^{-0.59}$.
%GPS второго типа связаны массово с HFP и локализованы в области
%более компактных размеров и высоких пиковых частот.
%Особый интерес могут представлять GPS низкой радиосветимости,
%которые массово для нашей выборки связаны с GPS первого типа и
%должны быть объектами высокочувствительных многочастотных обзоров.
%Среди обоих типов GPS присутствуют объекты малых угловых размеров
%с высокими значениями $\nu_{int} > 5 ГГц$ и низкой радиосветимостью ($10^{40-42}$ эрг/с).

The study and simulations of the morphological evolution of radio sources \cite{vris2009,an,collier} have provided several scenarios of the relation between the size and luminosity of extragalactic objects. The evolutionary differences are based on the initial conditions: properties of the environment and the behavior of the jet activity. As a result, the simple ``young-age'' scenario is not always evident, but is just one of the possible ones.

Some GPS studies  \cite{vris,mi} have found a deficit of GPS objects with peak frequencies lower than 1 GHz at high redshifts. In this study the deficit is evident despite the use of low-frequency measurements (see Fig. 11). A comparison of the ``$z$ -- $\nu_{int}$'' relation for our sample with the corresponding relation for a 110-object sample of the GLEAM survey \cite{callingham17} revealed their overall similarity at $z>2$: in both cases the number of objects with $\nu_{int}<1 GHz$ sharply decreases. The lack of large-scale components of synchrotron radiation during early Universe epochs is considered to be one of the possible causes of such deficit \cite{vris}.

In general,  type 1 GPS objects are less bright than type 2 objects and therefore the mean redshift of the former averaged over the sample is smaller %and their contribution to the ``$z$ -- $\nu_{int}$'' at high redshifts. 
On the other hand there is a sharp drop of the type 2 GPS objects with low-frequency peak at high redshifts. 

We found a statistical steepening of radio spectra with redshift increasing. Such correlation is expected for distant galaxies and has been found and discussed before \cite{dea1990,dea1,coppejans15,coppejans16}. A comparative analysis  of our sample and the sample of 108 USS (ultra-steep spectrum) galaxies \cite{rott} revealed a similar linear regression ``$z - \alpha_{above}$'', however, in the case of our sample we found no significant correlation between these quantities (Fig. 6). Considering that the sample in \cite{rott} consists of galaxies and our sample is a mix of galaxies and quasars, such result can be explained by the different morphologies of objects at relatively low redshifts. This is also apparent in Fig. 6a: we see mostly powerful GPS objects with increasing $z$. Subdividing our sample into small $dz = 0.2$ bins allowed us to reveal a trend for spectral steepening (the median values). The variety of object properties  at relatively low redshifts shows up as the large scatter of $\alpha_{med}$, values, whereas starting with $z > 2$ we observe a well-defined linear trend.
%According to our results the second type GPS sources are massively associated with the HFP objects and they are localized in the more compact sizes and high peak frequencies area.

The flux densities estimates of the hot spots on the Planck mission microwave background maps provides an information about the behavior of the GPS spectra at millimeter- and submillimeter-wave ranges in addition to the data available from the Planck catalog. Deriving two frequency spectra of sources in this range showed that part of the GPS objects can be classified as ``intermediate'', synchrotron-type objects, which, along with synchrotron radiation, may suggest the presence of a dust component. The rise of the spectra at high frequencies may be an indicator of both the presence of a signal from cold galactic dust on Planck maps and of the internal dust contribution of the sources. Taking into account that the spectral indices calculations based on estimated data, more accurate conclusions are impossible.

It has been correctly noted \cite{tinti2006} that GPS studies are often based on classic bright samples, which rely on selection according to the shape of the spectrum without taking into account the physical features. A conservative approach leads to the absence of common criteria such as a compact structure (<kpc), polarization and variability, the presence of relativistic effects. Measurements of these values are available for a limited number of candidates; thus the samples are initially heterogeneous.

It is necessary to note the selection effects that affect our results: the absence of systematical decimiter-waveband measurements during the intitial selection of GPS objects, the lack of ``broad'' approach while constructing the sample, because the selection was made at the same frequency of 5 GHz, and also the lack of measurements of faint GPS \cite{mi}. %It is evident that during the initial selection a fraction of objects was lost due to the lack of measurements and the inability to classify them as GPS objects. The number of such objects is  unknown, they are faint radio sources and there are no systematic measurements for them.
%GPS objects are also of practical interest as bright compact sources with the stationary radio emission, so-called flux standarts. 

\section{Results}
We obtained the following main results based on the GPS objects  monitoring with the RATAN-600 radio telescope in 2006--2017:
\begin{enumerate}
\item  the quasi-simultaneous measurements of flux densities at 1.1, 2.3, 4.8, 7.7/8.2, 11.2, and 21.7 GHz obtained with the RATAN-600 have been collected into the catalog; the broad band radio spectra have been constructed for the sample in the range of 0.072--857 GHz, based on RATAN-600 data with additional measurements from the GLEAM, TGSS and Planck surveys.
\item  we selected 164 GPS objects and candidates, 17 of them have been classified as GPS for the first time. The average value of the radio luminosity for GPS sample at 5 GHz is $10^{43-44} erg/s$. We confirm the relatively small proportion of GPS objects (1-2\%) among bright AGNs.
The sample contains GPS objects of first (``classic'') and second (``blazar'') type in the proportion of 60\% to 40\%. The statistical difference between the spectral parameters of two types GPS objects can reveal the heterogeneity of the physical conditions of the synchrotron emission formation in compact extragalactic objects.
\item The deficit of GPS objects with low peak frequencies (less than 1 GHz) at high redshifts ($z > 2$) have been confirmed. The statistical steepening of the GPS spectra with the redshit have been found in the considered cosmological epochs. At redshifts  $z \leq 1.5$ the contribution of the objects of different morphology is clearly noticeable, and at $z > 2$, the contribution of GPS objects with steeper spectra in optically thin emission regions increases.
\end{enumerate}

\begin{acknowledgments}
Observations with the RATAN-600 radio telescope are supported by the Ministry of Science 
and Higher Education of the Russian Federation (contract no. N007-03-283).
MGM and TVM are grateful for finance subsidized within the
government support of the Kazan (Volga region) Federal University
in order to improve their competitiveness among the world leading
research and educational centers.
\end{acknowledgments}

\clearpage
\newpage
\section*{Supplementary materials}

\begin{figure}[h]
\onelinecaptionsfalse
\centerline{
\includegraphics[angle=0,width=0.9\textwidth,clip]{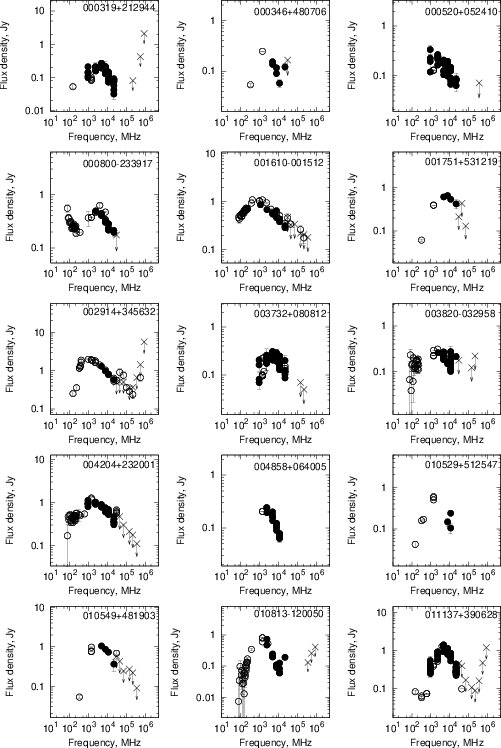} %3
}
\setcaptionmargin{0mm}
\captionstyle{normal}
\caption{Continuum radio spectra of GPS objects and candidates.
The RATAN-600 measurements in 2006-2017 are shown with filled circles, and data from the CATS database -- with empty circles, 
upper limits estimations of the flux densities from the Planck maps are shown with crosses.}
\label{fig3}
\end{figure}

\clearpage
\newpage

\begin{figure}[h]
%\onelinecaptionsfalse
\setcaptionmargin{5mm}
\onelinecaptionstrue
\centerline{
\includegraphics[angle=0,width=0.9\textwidth,clip]{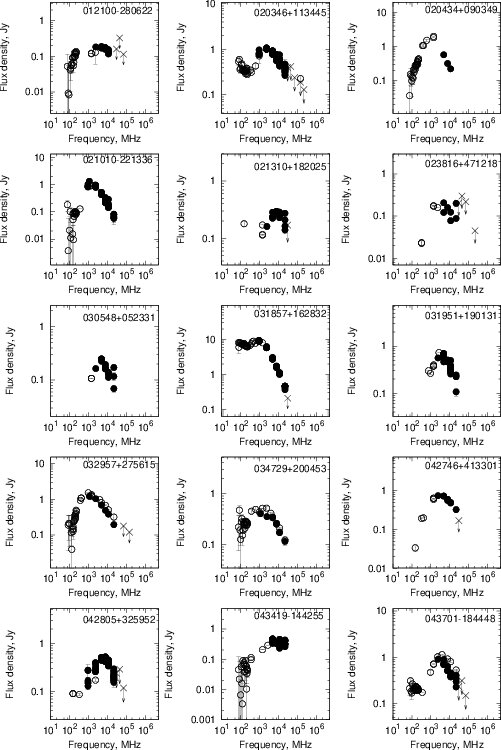} %4
}
\setcaptionmargin{0mm}
{Fig.~3. Continuation.}
\end{figure}

\clearpage
\newpage

\begin{figure}[h]
%\onelinecaptionsfalse
\setcaptionmargin{5mm}
\onelinecaptionstrue
\centerline{
\includegraphics[angle=0,width=0.9\textwidth,clip]{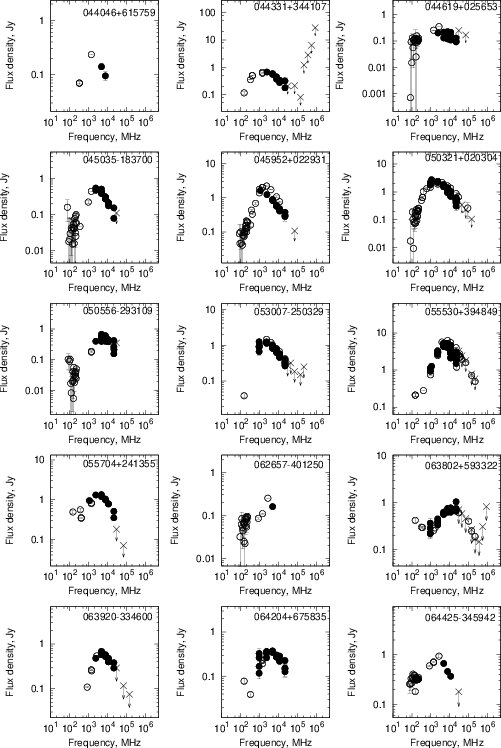} %5
}
\setcaptionmargin{0mm}
{Fig.~3. Continuation.}
\end{figure}

\clearpage
\newpage

\begin{figure}[h]
\setcaptionmargin{5mm}
\onelinecaptionstrue
\centerline{
\includegraphics[angle=0,width=0.9\textwidth,clip]{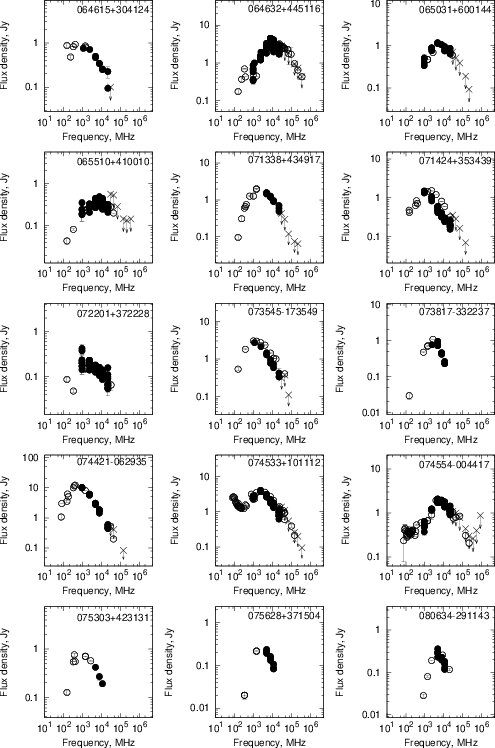} %6
}
\setcaptionmargin{0mm}
{Fig.~3. Continuation.}
\end{figure}

\clearpage
\newpage

\begin{figure}[h]
\setcaptionmargin{5mm}
\onelinecaptionstrue
\centerline{
\includegraphics[angle=0,width=0.9\textwidth,clip]{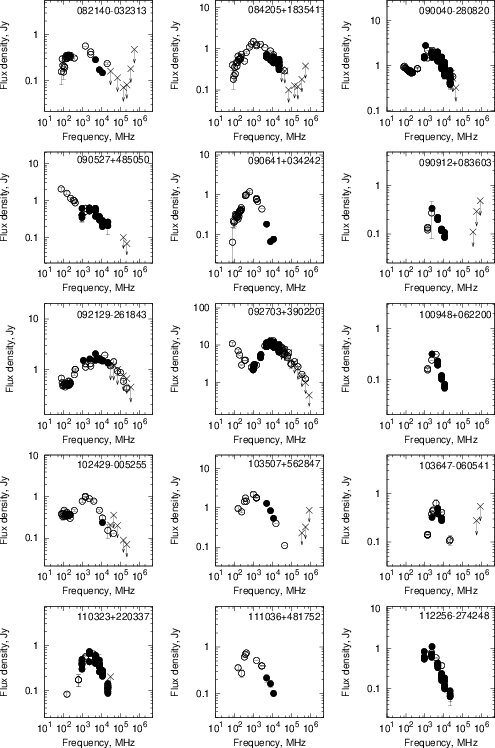} %7
}
\setcaptionmargin{0mm}
{Fig.~3. Continuation.}
\end{figure}

\clearpage
\newpage

\begin{figure}[h]
\setcaptionmargin{5mm}
\onelinecaptionstrue
\centerline{
\includegraphics[angle=0,width=0.9\textwidth,clip]{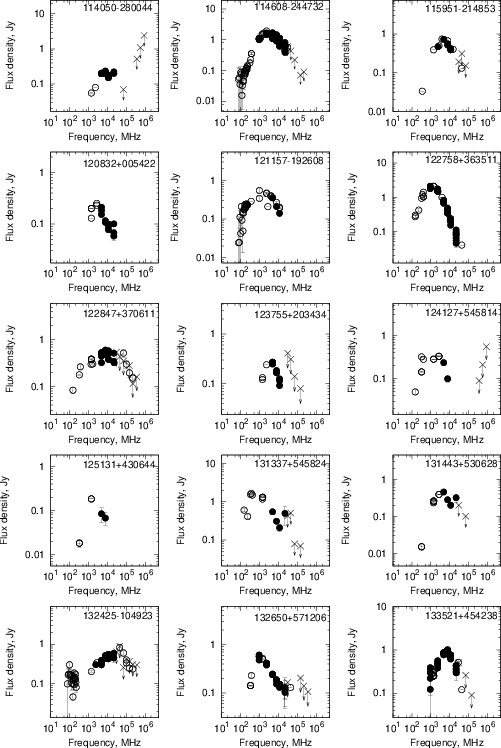}%8
}
\setcaptionmargin{0mm}
{Fig.~3. Continuation}
\end{figure}

\clearpage
\newpage

\begin{figure}[h]
\setcaptionmargin{5mm}
\onelinecaptionstrue
\centerline{
\includegraphics[angle=0,width=0.9\textwidth,clip]{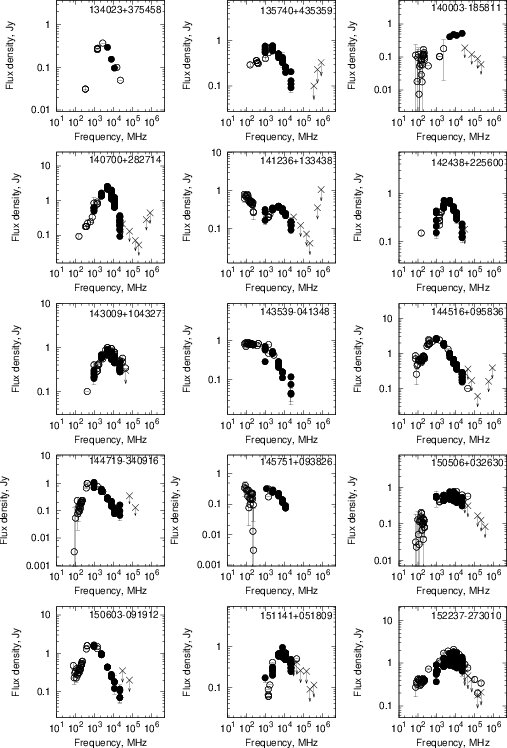} %9
}
\setcaptionmargin{0mm}
{Fig.~3. Continuation.}
\end{figure}

\begin{figure}[h]
\setcaptionmargin{5mm}
\onelinecaptionstrue
\centerline{
\includegraphics[angle=0,width=0.9\textwidth,clip]{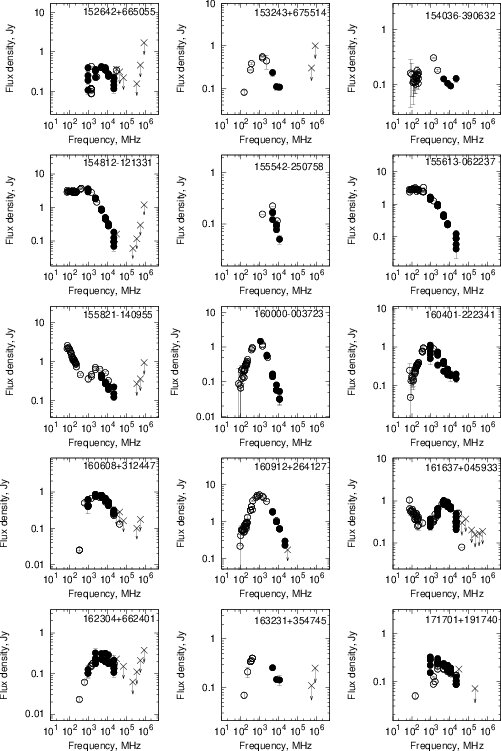}%10
}
\setcaptionmargin{0mm}
{Fig.~3. Continuation}
\end{figure}

\begin{figure}[h]
\setcaptionmargin{5mm}
\onelinecaptionstrue
\centerline{
\includegraphics[angle=0,width=0.9\textwidth,clip]{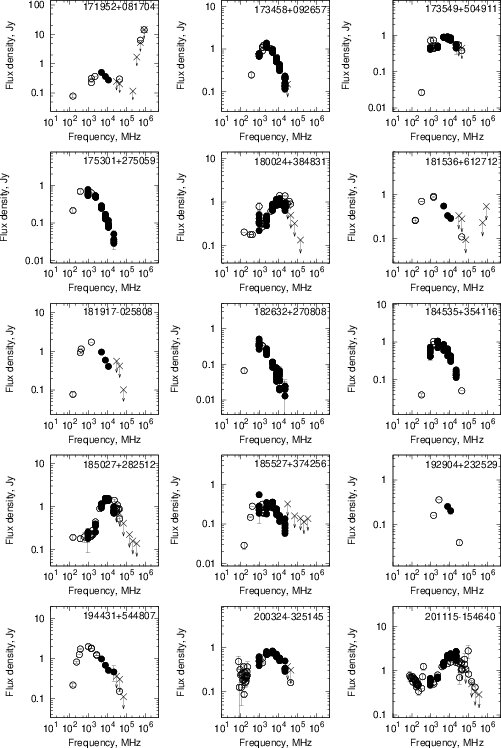} %11
}
\setcaptionmargin{0mm}
{Fig.~3. Continuation.}
\end{figure}

\begin{figure}[h]
\setcaptionmargin{5mm}
\onelinecaptionstrue
\centerline{
\includegraphics[angle=0,width=0.9\textwidth,clip]{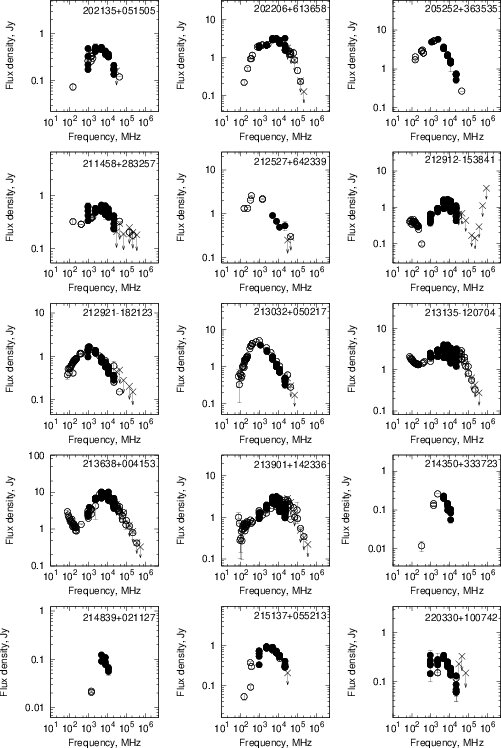} %12
}
\setcaptionmargin{0mm}
{Fig.~3. Continuation.}
\end{figure}

\begin{figure}[h]
\setcaptionmargin{5mm}
\onelinecaptionstrue
\centerline{
\includegraphics[angle=0,width=0.9\textwidth,clip]{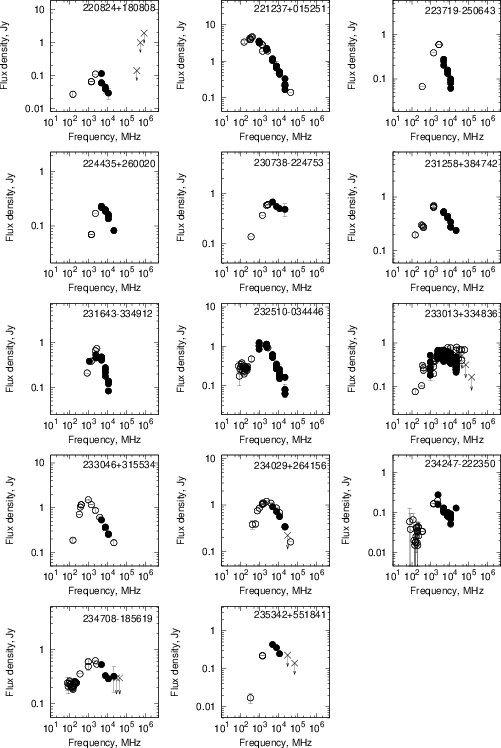} %13
}
\setcaptionmargin{0mm}
{Fig.~3. Continuation.}
\end{figure}

%\clearpage
%\newpage

\end{document}